\documentstyle[preprint,aps,eqsecnum]{revtex}

\def\beq#1{\begin{equation} \label{#1}}
\def\eeq{\end{equation}}

\begin{document}
{
\tighten
%\preprint {\vbox{
% \hbox{WIS-9y/XX/XXX-PH}
% \hbox{TAUP YYYY-98}
% \hbox{hep-ph/XXXXXX}
% \hbox{}
%}}

\title{A useful approximate isospin equality for charmless strange B Decays }

\author{Harry J. Lipkin\,$^{a,b}$}

\address{ \vbox{\vskip 0.truecm}
  $^a\;$Department of Particle Physics \\
  Weizmann Institute of Science, Rehovot 76100, Israel \\
\vbox{\vskip 0.truecm}
$^b\;$School of Physics and Astronomy \\
Raymond and Beverly Sackler Faculty of Exact Sciences \\
Tel Aviv University, Tel Aviv, Israel}

\maketitle

\begin{abstract}

A useful inequality is obtained if charmless strange B decays are assumed to be
dominated by the gluonic penguin diagram and the tree diagram contribution is
small but not negligible. The penguin-tree interference contribution which is
linear in the tree amplitude is included but the direct tree contribution which
is quadratic is neglected.

\end{abstract}
} % end tighten

It is now believed that charmless strange B decays are dominated by the gluonic
penguin diagram. If all other diagrams are negligible, the branching ratios for
sets of decays to states in the same isospin multiplets are uniquely related by
isospin because the gluonic penguin leads to a pure I=1/2 final state, and there
is no simple mechanism in the standard model that can give CP violation.

If the tree diagram produced by the $\bar b \rightarrow \bar u u \bar s$
transition at the quark level also contributes, the isospin analysis becomes
non-trivial, much more interesting and CP violation can be observed. This case
has been considered in great detail by Nir and Quinn\cite{PBPENG}.

Our purpose here is to point out and discuss an intermediate case, where the
amplitude from the tree diagram is not negligible but still sufficiently small
by comparison with the gluonic penguin that its contribution to branching ratios
need be considered only to first order, and higher order contributions can be
neglected. An interesting approximate sum rule is obtained which can check the
validity of this approximation and guide the search for CP violation.

We consider here the  $ B \rightarrow K \pi $ decays. Exactly the same
considerations hold for all similar decays into a strange meson and an isovector
nonstrange meson.

The gluonic penguin diagram leads to a pure isospin 1/2 state. Its contributions
to all $ B \rightarrow K \pi $ amplitudes, denoted by $P$ are simply related by
isospin.
$$ P(B^+ \rightarrow K^+ \pi^o) =
- P (B^o \rightarrow K^o \pi^o) =
{1 \over \sqrt 2} \cdot P (B^o \rightarrow K^+ \pi^-) =
{1 \over \sqrt 2} \cdot P (B^+ \rightarrow K^o \pi^+) \equiv P
\eqno(1)                                          $$
The tree diagram has two independent contributions corresponding to the
color-favored and color-suppressed couplings of the final quark-antiquark pairs.
Their amplitudes, denoted by $T_f$ and $T_s$ are also simply related by isospin.
$$ {1 \over \sqrt 2} \cdot T_f (B^o \rightarrow K^+ \pi^-) =
T_f (B^+ \rightarrow K^+ \pi^o)
\equiv T_f
\eqno(2a)                      $$
$$ T_s (B^+ \rightarrow K^+ \pi^o) =
T_s (B^o \rightarrow K^o \pi^o)
\equiv T_s
\eqno(2b)                  $$
In the approximation where we consider the contributions of the tree amplitudes
only to first order, we obtain:
$$ BR (B^o \rightarrow K^+ \pi^-) \approx 2 P^2 + 4 P \cdot T_f
\eqno(3a)                                          $$
$$ BR (B^o \rightarrow K^o \pi^o) \approx P^2 - 2  P \cdot T_s
\eqno(3b)                                          $$
$$ BR (B^+ \rightarrow K^o \pi^+) \approx 2 P^2
\eqno(3c)                                          $$
$$ BR (B^+ \rightarrow K^+ \pi^o) \approx P^2 + 2 P \cdot T_f
+ 2 P \cdot T_s
\eqno(3d)                                          $$
This leads to the approximate equality
$$ BR (B^o \rightarrow K^+ \pi^-) -2 BR
(B^o \rightarrow K^o \pi^o) \approx
2 BR (B^+ \rightarrow K^+ \pi^o) - BR (B^+ \rightarrow K^o \pi^+) \approx $$
$$ \approx 4 P \cdot T_f + 4 P \cdot T_s =
4 P(T_f + T_s)\cdot \rm cos (\phi_P - \phi_T - \phi_S)
\eqno(4a)                                          $$
where $\phi_P$ and $\phi_T$ are the weak phases respectively of the penguin and
tree amplitudes and $\phi_S$ is the strong phase difference between the two.
Note that both the left-hand and right hand sides vanish independently for
any transition like the pure gluonic penguin that leads to a pure I=1/2 state.
The relation (4) therefore is due entirely to interference between the
penguin I=1/2 amplitude and the I=3/2 component of the tree amplitude.
This immediately leads to the correspnding expression for the charge-conjugate
decays,
$$ BR (\bar B^o \rightarrow K^- \pi^+) -2 BR
(\bar B^o \rightarrow \bar K^o \pi^o) \approx
2 BR (B^- \rightarrow K^- \pi^o) - BR (B^- \rightarrow \bar K^o \pi^-) \approx
$$
$$
4 P(T_f + T_s)\cdot \rm cos (\phi_P - \phi_T + \phi_S)
\eqno(4b)                                          $$
The direct CP violation is seen to be given by
$${{2 BR(B^- \rightarrow K^- \pi^o) - 2 BR (B^+ \rightarrow K^+ \pi^o) -
BR (B^- \rightarrow \bar K^o \pi^-) + BR (B^+ \rightarrow K^o \pi^+)} \over
{2 BR(B^- \rightarrow K^- \pi^o) + 2 BR (B^+ \rightarrow K^+ \pi^o) -
BR (B^- \rightarrow \bar K^o \pi^-) - BR (B^+ \rightarrow K^o \pi^+)}}
\approx $$
$$\approx
 \rm tan (\phi_P - \phi_T) \rm tan ( \phi_S)
\eqno(5)                                          $$
The same equality (4) is easily seen in the formalism used by Nir and
Quinn\cite{PBPENG} using their three amplitudes denoted by $U$, $V$ and $W$.
The $W$ amplitude is the penguin amplitude and the the two tree amplitudes
$U$ and $V$ are linear combinations of our $T_f$ and $T_s$ amplitudes,
defined by isospin properties rather than quark diagrams.
$$ BR (B^o \rightarrow K^+ \pi^-) -2 BR (B^o \rightarrow K^o \pi^o) =
-2W\cdot(U+V) + 2(V^2 - U^2) \approx -2W\cdot(U+V)
\eqno(6a)                                          $$
$$ 2 BR (B^+ \rightarrow K^+ \pi^o) - BR (B^+ \rightarrow K^o \pi^+)
= -2W\cdot(U+V) -  2(V^2 - U^2) \approx -2W\cdot(U+V)
\eqno(6b)                                          $$
It is interesting to note that here also one sees the apparent miracle that
the $same$ linear combination of the two tree amplitudes, $U+V$ appears in
both expressions (6a) and (6b) and that the approximate equality follows
from the approximation that quadratic terms in U and V are negligible in
comparison with the product of the linear terms and the dominant penguin
amplitude W.
$$ (V^2 - U^2) <<  W\cdot(U+V)    \eqno(7)               $$

The relations (3) neglect the contribution of annihilation diagrams and
diagrams in which a flavor-changing final state interaction like charge exchange
follows the weak tree diagram. However, both the annihilation diagram and the
charge exchange diagram which proceeds via the quark annihilation and pair
creation transition $u \bar u \rightarrow $ gluons $\rightarrow d \bar d$ lead
to pure I=1/2 final states and their contributions cancel in the sum rule (4).
The contributions of electroweak penguin amplitudes are not taken into account.

\acknowledgments
It is a pleasure to thank Yosef Nir for helpful discussions and comments. This
work was partially supported by the German-Israeli Foundation for Scientific
Research and Development (GIF).

{
\tighten

}

\end{document}